\documentclass{elsart}

\usepackage{psfig}

\usepackage{natbib}

\begin{document}

\runauthor{Cicero, Caesar and Vergil}

% -------------------------------------------------------------------------

\begin{frontmatter}

\title{Discovery of a Narrow-Line Seyfert 1 galaxy} 

\author[TIFR]{K. P. Singh}
\address[TIFR]{Department of Astronomy \& Astrophysics, Tata Institute of Fundamental Research, Mumbai, India}
\author[BIRM]{L. R. Jones}
\address[BIRM]{Department of Physics \& Astronomy, University of Birmingham, U.K.}

\begin{abstract}
We have discovered a Narrow-Line Seyfert 1 (NLS1) galaxy and identified 
it with an ultrasoft X-ray source {\bf (RX~215319-1514)} detected
with $ROSAT$.  We present its X-ray and optical spectral characteristics.  
Its redshift is found to be 0.0778$\pm$0.0002.  Its optical spectrum 
shows fairly strong Fe II emission  lines, and its X-ray spectrum has 
an extremely steep power-law index of $\Gamma$ $\geq$4, making it an 
extremely interesting example of this class of AGNs.  No low-energy 
absorption or variability is observed.  The soft X-ray (0.1--2.0 keV) 
luminosity is estimated to be 3.4--12.5$\times$10$^{43}$ ergs s$^{-1}$, 
depending on the spectral model.
\end{abstract}


\end{frontmatter}

% -------------------------------------------------------------------------

\section{Introduction}

Narrow-line Seyfert 1 (NLS1) galaxies with narrow 
(FWHM= 500--2000 km s$^{-1}$) emission lines of hydrogen in their
optical/UV spectrum tend to have the steepest soft X-ray spectra in the
$ROSAT$ band of 0.1--2.0 keV, and form a distinct class of active
galactic nuclei (see Boller et al. 1996 and references therein).  
Catalogues of ultra-soft X-ray sources are, therefore, very useful to 
find such objects.  While carrying out optical spectroscopy of candidate
objects in the positional error circles of such
sources from the catalogue given by  Singh et al. (1995), we 
have found a new narrow-line Seyfert 1 (NLS1) galaxy. The galaxy is a 
strong candidate for being a counterpart to the ultra-soft X-ray 
source -- WGA~J2153.3-1514 (RXJ~2153.3-1514). A brief summary of its 
optical and X-ray properties is presented below.

\section{X-ray and Optical Observations}

Analysis of archival $ROSAT$ PSPC data in the vicinity of WGA~J2153.3-1514 
shows two bright optical objects near the X-ray peak -- a 14.7 mag 
galaxy and a 11.72 mag star (see Figure 1).
The X-ray beam is broadened, and somewhat distorted, due to 
the large off-axis angle (41 arcmin) of the source. 

\begin{figure}[htb]
\centerline{\psfig{figure=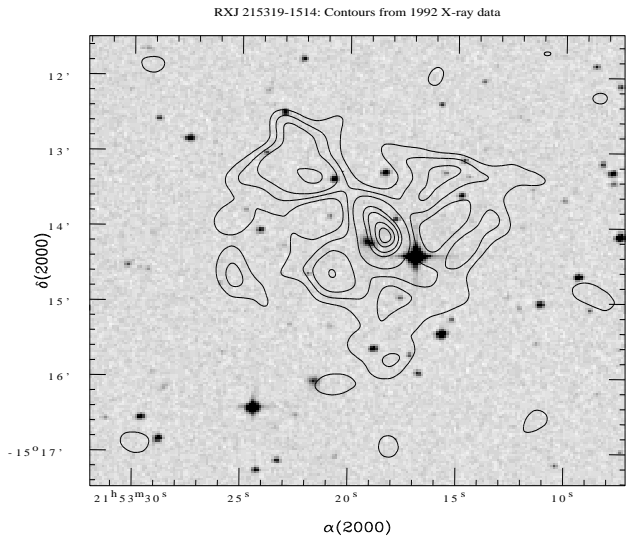,height=3.9truein,width=3.5truein,angle=0}}
\caption{Overlay of X-ray contour map from $ROSAT$ PSPC observations on optical image from digitised sky survey.}
\end{figure}

Optical spectra were taken for both the objects at the Lick 3m telescope
[Slit width = 2.5 arcsec; Grism for blue and a grating for red spectrum
both of 600 lines/mm gave a resolution of 6.5 \AA~(FWHM)].
The spectrum of the galaxy is shown in Figure 2  after
smoothing.  The spectral parameters for the emission lines are given
in Table 1.  Spectral features are consistent with the classification as
a NLS1 galaxy (Osterbrock \& Pogge 1985; Goodrich 1989).
The star shows a spectrum of a late-A to an early-F type, and no
chromospheric emission or emission lines characterisitic of an accreting
cataclysmic variable are detected.  These properties and the ratio of
X-ray flux (F$_x$) to optical flux (F$_v$), effectively rule out the
star as being the candidate for the X-ray source.

\vspace{3.0cm}
\centerline {\bf Table 1. Optical Emission Line Parameters of the NLS1 galaxy}
\begin{tabular}{lccc}
 Emission   & Equivalent    & Measured     & Intrinsic \\
  Line      & Width (\AA)   &  FWHM (\AA)  & FWHM (km s$^{-1}$)\\[10pt]
NII 6549 \AA & 1.0$\pm$0.3    & 1.1$\pm$0.5  & - \\
H$\alpha$ & 40.8$\pm$1.0    & 22.3$\pm$1.2 & 880$\pm$50 \\
NII 6583 \AA & 9.8$\pm$0.6    & 7.8$\pm$0.6  & 180$\pm$25 \\ 
\end{tabular}

\begin{figure}[htb]
\centerline{\psfig{figure=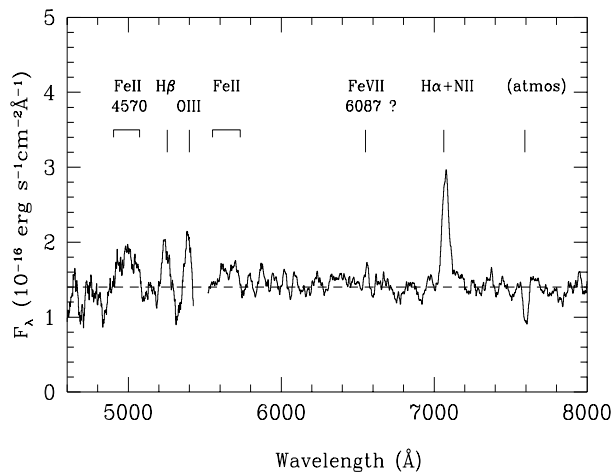,height=3.9truein,width=3.5truein,angle=0}}
\caption{Optical spectrum (smoothed to show Fe II features) of the active galaxy.}
\end{figure}

X-ray observations from 2 epochs -- 1992 and 1993, analysed separately as
well as jointly for their spectral content, show a very steep 
spectrum ($\Gamma$ $\geq$4 for power-law model).
Spectral parameters for the power-law and other models are given in Table 2.
No variability in the X-ray flux or spectral parameters is detected.
Presence of any low-energy absorption local to the source is also not detected.
Galactic N$_{\rm H}$ in this direction is 4$\times$10$^{20}$ cm$^{-2}$, 
based on 21~cm radio observations.

\vspace{3.5cm}
\centerline {\bf Table 2. X-ray Spectral Parameters}
\begin{tabular}{lccccc} 
  Model & $\Gamma$/kT(eV) & N$_{\rm H}$ & F$_x$ & S$_x$ & L$_x$  \\
        &        &  10$^{20}$  & 10$^{-14}$ergs & 10$^{-12}$ergs & 10$^{43}$ergs s$^{-1}$ \\
        &        &  cm$^{-2}$  & cm$^{-2}$s$^{-1}$ & cm$^{-2}$s$^{-1}$ & (z=0.0778) \\
        &        &             &         &         & (H$_o$=75;q$_o$=0)\\[10pt]  Power-law &  6.4$^{+5}_{-1.5}$ & 4$^{+3.9}_{-1.4}$ & 9.5 & 10 & 12.5 \\
  Black-body  & 30$^{+9}_{-9}$ & 4 & 7.5 & 2.7 & 3.4 \\
  MEKAL & 54$^{+11}_{-12}$ & 4 & 14 & 5.3 & 6.6 \\ 
\end{tabular}

Steep X-ray spectra, as are observed here,  are usually associated with 
NLS1 galaxies (Brandt, Pounds, \& Fink 1995; Boller, Brandt, \& Fink 1996;
Laor et al. 1997).
The position of this galaxy on the $\Gamma$-H$\beta$ line width plane
(Fig. 8 of Boller et al (1996)) is consistent with an extrapolation
based on the other NLS1 galaxies.
For further details, please refer to Singh \& Jones (2000).

Higher resolution X-ray observations with more sensitive satellites 
like $Chandra$ and $XMM-Newton$, as well as higher resolution optical/UV
spectra with better signal-to-noise ratio are required to fully understand
the nature of the AGN in this galaxy.

% -------------------------------------------------------------------------

% -------------------------------------------------------------------------

\end{document}